\documentclass[journal]{IEEEtran}

\usepackage{cite}

\usepackage{algorithmic}
\usepackage{algorithm2e}
\usepackage{caption}

\usepackage{color}
\usepackage{placeins}
\usepackage{float}
\usepackage{tabularx,colortbl}
\usepackage{multirow}

\usepackage{subfigure}

\usepackage[cmex10]{amsmath}
\usepackage[pdftex]{graphicx}
\usepackage{amsfonts}

\usepackage{xcolor}

\usepackage{array}

\newcolumntype{L}[1]{>{\raggedright\let\newline\\\arraybackslash\hspace{0pt}}m{#1}}
\newcolumntype{C}[1]{>{\centering\let\newline\\\arraybackslash\hspace{0pt}}m{#1}}
\newcolumntype{R}[1]{>{\raggedleft\let\newline\\\arraybackslash\hspace{0pt}}m{#1}}

\newcommand{\mbold}[1]{\mathbf{#1}}

\newcommand{\mvar}[1]{\boldsymbol{#1}}

\begin{document}
%
\title{Generalized Optimization of High Capacity Compressive Imaging Systems}

\author{\IEEEauthorblockN{Richard Obermeier and Jose~Angel~Martinez-Lorenzo}}



\maketitle

\begin{abstract}
One of the greatest challenges in applying compressive sensing (CS) signal processing techniques to electromagnetic imaging applications is designing a sensing matrix that has good reconstruction capabilities. Compressive reflector antennas (CRA) are a class of antennas that have been shown to provide enhanced image reconstruction performance over traditional reflector antennas (TRA) when CS techniques are employed. In this paper, we present a unified CRA design method, which considers both the sensing capacity and efficiency of the antenna, and can be used for both compressive imaging and multiple-input multiple-output (MIMO) communication applications. The unified design method is assessed for a CRA configuration in which dielectric scatterers are added to the surface of a TRA. The design results demonstrate the ability of the unified design method to enhance the CS reconstruction capabilities of the CRA.
\end{abstract}


%
\IEEEpeerreviewmaketitle

\vspace{7pt}
\section{Introduction}
\label{sec:intro}

Compressive reflector antennas (CRA) \cite{Martinez2015a,Heredia-Juesas2015,obermeier2016model} are a class of antennas specifically designed for high capacity sensing and imaging applications. CRA's operate in a manner similar to that of the coded apertures utilized in optical imaging applications \cite{busboom1997coded,de2009sub,marcia2008compressive}: by introducing scatterers to the surface of a traditional reflector antenna (TRA), the compressive antenna encodes a pseudo-random phase front on the scattered electric field. It was shown in \cite{Martinez2015a} that, by modifying the encoded wavefront from measurement to measurement, compressive sensing (CS) techniques \cite{Candes2006,Candes2006a,Donoho2006} can be employed in imaging applications with improved performance over the traditional reflector antenna.

In \cite{obermeier2016model}, a numerical design method for compressive antennas was developed based on the notion that enhancing the capacity of the antenna's Green's functions should also enhance the capacity of the sensing matrix in electromagnetic imaging applications. While this notion holds true for monostatic imaging applications, it does not necessarily hold true for multi-static imaging applications because it does not properly consider the coupling between the transmitting and receiving antennas. The simplified method described in \cite{obermeier2016model} also utilized a full-wave numerical model based on finite differences in the frequency domain \cite{Rappaport2001}, which is not very well suited for practical imaging applications due to its large computational complexity.

In this paper, we extend the previous work on compressive antenna design in two ways. First, we introduce a unified design method that properly considers the coupling between the transmitting and receiving antennas. This method also considers the efficiency of the antenna (i.e. how much energy is radiated into the imaging region), and can be used for both compressive imaging and multiple-input multiple-output (MIMO) communication applications. Second, we formulate an instance of the unified design method using an efficient forward model known as the modified equivalent current approximation (MECA) \cite{meana2010wave,gutierrez2011high}, which is well suited for large three-dimensional imaging applications.

The remainder of this paper is organized as follows. In Section \ref{sec:motivation}, we describe the motivation for using sensing capacity as a design metric in compressive imaging applications. In Section \ref{sec:sensing_model}, we describe the sensing model and introduce the unified antenna design optimization problem. In Section \ref{sec:meca}, we describe how MECA can be used in the unified design method to design compressive antennas with dielectric scatterers. In Section \ref{sec:results}, we present design results for realistic compressive imaging and communication problems. Finally, we conclude the paper in Section \ref{sec:conclusion}.

\vspace{7pt}
\section{Motivation}
\label{sec:motivation}
Consider a general linear system, in which a set of noisy measurements $\mbold{y} \in \mathbb{C}^M$ of the object of interest $\mvar{x} \in \mathbb{C}^N$ are obtained via the relationship $\mbold{y} = \mbold{A} \mvar{x} + \mbold{n}$, where $\mbold{A} \in \mathbb{C}^{M \times N}$ is called the sensing matrix. When $M < N$, there are an infinite number of solutions $\mbold{x}$ satisfying $\mbold{y} = \mbold{A}\mvar{x}$, and so regularization techniques must be employed in order to generate a unique solution. A particularly interesting scenario arises when the vector $\mvar{x}$ is sparse, that is the number of non-zero elements $S$ in $\mvar{x}$ is much smaller than the total number of elements $N$. In this case, compressive sensing theory \cite{Candes2006,Candes2006a,Donoho2006} establishes that the sparse vector can be stably recovered from an incomplete set of measurements as the solution to the following convex optimization program:
\begin{alignat}{3}
& \underset{\mvar{x}}{\text{minimize }}~~ & &\|\mvar{x}\|_{\ell_1} \label{eq:ell1norm-1} \\
&\text{subject to } &  &\|\mbold{A}\mvar{x} - \mbold{y}\|_{\ell_2} \le \eta \nonumber
\end{alignat}
Stable reconstruction is only guaranteed when the sensing matrix $\mbold{A}$ is ``well behaved'' according to a suitable measure. The Restricted Isometry Property (RIP), arguably the most popular measure, uses the concept of restricted isometry constants in order to establish the tightest performance guarantees currently known. For a fixed sparsity level $S$, the restricted isometry constant $\delta_S$ is the smallest positive constant such that
 \begin{align}
 (1-\delta_S)\|\mbold{x}\|_{\ell_2}^2 \le \|\mbold{Ax}\|_{\ell_2}^2 \le (1+\delta_S)\|\mbold{x}\|_{\ell_2}^2 \label{eq:ric}
 \end{align}
is satisfied for all vectors with $\|\mbold{x}\|_{\ell_0} \le S$, where the ``$\ell_0-$norm'' measures the number of non-zero elements in the vector. In other words, the restricted isometry constant $\delta_S$ establishes bounds for the singular values of submatrices obtained by selecting any $S$ columns from the complete sensing matrix $\mbold{A}$. Generally speaking, the RIP requires the values of $\delta_S$ to be small.

Compressive sensing can also be considered from the perspective of information theory. In this framework, the $\epsilon$-capacity determines the amount of information that can be transmitted through the linear mapping $\mbold{y} = \mbold{A} \mvar{x}$ within an uncertainty level $\epsilon$ \cite{migliore2008electromagnetics,migliore2011}, and is defined as follows:
\begin{align}
H_\epsilon(\mbold{A}) =  \frac{1}{2}\log_2\left(\frac{\det \mbold{A}\mbold{A}^H}{\epsilon^2}\right) = \sum_{m=1}^M\log_2\left(\frac{\sigma_m}{\epsilon}\right) \label{eq:epscapacity}
\end{align}
where $\sigma_m$ are the singular values of $\mbold{A}$. We refer to the $\epsilon$-capacity as the sensing capacity, or just the capacity.
In \cite{obermeier2016model}, the sensing capacity for the complete matrix $A$ was shown to establish the following lower bound for the restricted isometry constants $\delta_S$:
\begin{align}
\delta_S \ge 1 - 2^{2H_1(\mbold{A})} = 1 - \prod_{m=1}^M \sigma_m^2
\end{align}
By increasing the sensing capacity, the lower bound for the restricted isometry constants is improved, thereby improving the likelihood that compressive sensing techniques will be successful. 


\vspace{7pt}
\section{Unified Design Problem}
\label{sec:sensing_model}
Consider an electromagnetic sensing system that utilizes several transmitting and receiving antennas. The fields radiated by the transmitting antenna on the $m$-th measurement can be denoted by the vector $\mbold{g}_{t,i,m} \in \mathbb{C}^{N}$, where $i=1,2,3$ denotes the component of the electric field and $N$ denotes the number of points in the imaging region. Similarly, the radiation pattern of the receiving antenna on the $m$-th measurement can be denoted by the vector $\mbold{g}_{r,i,m} \in \mathbb{C}^{N}$. By concatenating the radiated fields for all $M$ measurements, we can formulate the field matrices $\mbold{G}_{t,i}, \mbold{G}_{r,i} \in \mathbb{C}^{M \times N}$. According to the reciprocity theorem, the sensing matrix $\mbold{A} \in \mathbb{C}^{M \times N}$ can be expressed as $\mbold{A} = \sum_{i=1}^3 \zeta \left(\mbold{G}_{r,} \odot \mbold{G}_{t,i}\right)$, where $\zeta$ is a positive scaling constant and $\odot$ is the Hadamard (element-wise) product. This is an instance of the Born approximation.

Suppose that the radiated fields can be expressed as a function of design variables $\mvar{x} \in \mathbb{R}^P$. The unified design problem seeks the optimal solution to the following non-convex optimization program: 
\begin{alignat}{3}
& \underset{\mvar{x}}{\text{maximize }}~~ & & \alpha_A\log\det\left(\mbold{A}(\mvar{x})\mbold{A}^H(\mvar{x})  + \beta_A \mbold{I}_M\right) \label{eq:unified_opt} \\
& & & + \sum_{i=1}^3 \alpha_{t,i} \log\det\left(\mbold{G}_{t,i}(\mvar{x})\mbold{G}^H_{t,i}(\mvar{x}) + \beta_{t,i} \mbold{I}_M\right) \nonumber \\
& & & + \sum_{i=1}^3  \alpha_{r,i}\log\det\left(\mbold{G}_{r,i}(\mvar{x})\mbold{G}^H_{r,i}(\mvar{x})  + \beta_{r,i} \mbold{I}_M \right) \nonumber \\
& & & + \sum_{m=1}^M \mbold{a}_{m}(\mvar{x})\mbold{\Lambda}_{A,m} \mbold{a}^H_{m}(\mvar{x}) \nonumber  \\
& & & + \sum_{m=1}^M \sum_{i=1}^3 \mbold{g}_{t,i,m}(\mvar{x})\mbold{\Lambda}_{t,i,m}\mbold{g}^H_{t,i,m}(\mvar{x}) \nonumber \\ 
& & & + \sum_{m=1}^M \sum_{i=1}^3 \mbold{g}_{r,i,m}(\mvar{x})\mbold{\Lambda}_{r,i,m} \mbold{g}^H_{r,i,m}(\mvar{x}) \nonumber \\
&\text{subject to } &  & \mvar{x}\in Q_p \nonumber
\end{alignat}
where $\mbold{a}_{m}(\mvar{x}) = \sum_{i=1}^3 \zeta \left(\mbold{g}_{t,i,m}(\mvar{x}) \odot \mbold{g}_{r,i,m}(\mvar{x})\right)$ and $Q_p$ is the feasible set for the design variables. The coefficients $\alpha_{t,i}$, $\alpha_{r,i}$, and $\alpha_{A}$ are positive weights associated with each of the sensing capacity terms, $\mbold{\Lambda}_{t,i,m}$, $\mbold{\Lambda}_{r,i,m}$, and $\mbold{\Lambda}_{A,m}$ are diagonal matrices that determine how the energy should be distributed in the imaging region, and $\beta_{t,i}$, $\beta_{r,i}$, and $\beta_{A}$ are constants that ensure that the matrices input to $\det$ have full rank. These parameters can be tuned from one application to another in order to achieve the desired effects. For example, one might set $\alpha_{A,m}=1$ and all other parameters to zero when designing an antenna for compressive imaging applications. When designing a receiving antenna, one might set $\alpha_{r,i} > 0$ and all other parameters to zero. More complex scenarios might consider the $\mbold{\Lambda}$ matrices, whose elements can be positive or negative. Positive elements are associated with voxels where we want to deposit energy, negative elements are associated with voxels that we would like to null, and zero elements are associated with voxels that we are indifferent to.

If $\mbold{g}_{t,i,m}(\mvar{x})$ and $\mbold{g}_{r,i,m}(\mvar{x})$ are differentiable, then the objective function of Eq. \ref{eq:unified_opt} is differentiable. Furthermore, if the feasible set $Q_p$ is convex and has an easy to compute proximal operator, then Eq. \ref{eq:unified_opt} can be solved using proximal gradient techniques. To start, we note that the partial derivatives $\frac{\partial \mbold{a}_m(\mvar{x})}{\partial x_p}$ can be computed from $\frac{\partial \mbold{g}_{t,i,m}(\mvar{x})}{\partial x_p}$ and $\frac{\partial \mbold{g}_{r,i,m}(\mvar{x})}{\partial x_p}$ as follows:
\begin{align}
\frac{\partial \mbold{a}_m(\mvar{x})}{\partial x_p} &= \sum_{i=1}^3 \zeta\left(\frac{\partial \mbold{g}_{t,i,m}(\mvar{x})}{\partial x_p} \odot \mbold{g}_{r,i,m}(\mvar{x})\right) \\
&+ \sum_{i=1}^3 \zeta\left(\mbold{g}_{t,i,m}(\mvar{x}) \odot \frac{\partial \mbold{g}_{r,i,m}(\mvar{x})}{\partial x_p} \right) \nonumber
\end{align}
The partial derivatives $\frac{\partial \mbold{G}_{t,i}(\mvar{x})}{\partial x_p}$, $\frac{\partial \mbold{G}_{r,i}(\mvar{x})}{\partial x_p}$, and $\frac{\partial \mbold{A}(\mvar{x})}{\partial x_p}$ can then be computed by concatenating the partial derivatives $\frac{\partial \mbold{g}_{t,i,m}(\mvar{x})}{\partial x_p}$, $\frac{\partial \mbold{g}_{r,i,m}(\mvar{x})}{\partial x_p}$, and $\frac{\partial \mbold{a}_m(\mvar{x})}{\partial x_p}$, respectively. From here, the partial derivatives required for Eq. \ref{eq:unified_opt} can be expressed in terms of the partial derivatives of ${f}_1(\mvar{x}) = \log\det(\mbold{F}(\mvar{x}))$ and ${f}_2(\mvar{x}) = \mbold{f}(\mvar{x}) \mbold{\Lambda} \mbold{f}^H(\mvar{x})$. These derivatives have the following closed-form solutions:
\begin{align}
\frac{\partial {f}_1(\mvar{x})}{\partial x_p} &= \operatorname{tr}\left(\mbold{F}^{-1}(\mvar{x}) \frac{\partial \mbold{F}(\mvar{x})}{\partial x_p}\right) \\
\frac{\partial {f}_2(\mvar{x})}{\partial x_p} &=\frac{\partial \mbold{f}(\mvar{x})}{\partial x_p} \mbold{\Lambda} \mbold{f}(\mvar{x})^H + \mbold{f}(\mvar{x}) \mbold{\Lambda} \frac{\partial \mbold{f}(\mvar{x})^H}{\partial x_p}
\end{align}
where $\operatorname{tr}(\cdot)$ is the trace of a matrix.

The proximal gradient method solves the optimization problem of Eq. \ref{eq:unified_opt} using the following steps. First, the gradient $\mbold{h}_k$ is computed at the current iteration point $\mbold{x}^{(k)}$. Second, a proximal step is computed along the search direction $\mbold{h}_k$ to compute the next iteration point $\mbold{x}^{(k+1)}$. Formally, this point can be expressed as the solution to the following convex optimization problem:
\begin{align}
\underset{\mvar{x}}{\text{minimize }}  I_{Q_p}(\mvar{x}) + \frac{1}{2\kappa}\|\mvar{x} - \left(\mbold{x}^{(k+1)} - \kappa \mbold{h}_k\right)\|_{\ell_2}^2 \label{eq:gradprox}
\end{align}
where $I_{Q_p}(\mvar{x})$ is the indicator function for the feasible set. This is nothing more than the proximal operator for the feasible set. In practice, a line search method like the one presented in \cite{Beck2009a} is used to solve Eq. \ref{eq:gradprox}.

\vspace{7pt}
\section{CRA Design using MECA}
\label{sec:meca}
\subsection{General Formulation}
MECA \cite{meana2010wave,gutierrez2011high} is an efficient extension to the physical optics (PO) method for penetrable and non-metallic objects. In MECA, scattering elements are discretized into triangular facets on which equivalent electric and magnetic currents are computed. By treating these currents as sources radiating in the background medium whose Green's functions are known, MECA can compute the electric and magnetic fields at any position. Formally, the equivalent currents on the triangular facets are expressed as
\begin{alignat}{3}
& \mbold{J} = \mbold{\hat{n}} \times \mbold{H}_t\bigg|_S &=&~ \frac{1}{\eta_1} E_{TE_i}\cos(\theta_i)\left(1 - \Gamma_{TE}\right)\mbold{\hat{e}}_{TE} \label{eq:Jmeca} \\
& &+&~ \frac{1}{\eta_1} E_{TM_i}\left(1 - \Gamma_{TM}\right)(\mbold{\mbold{\hat{n}}\times\hat{e}}_{TE})\bigg|_S \nonumber \\
& \mbold{M} = -\mbold{\hat{n}} \times \mbold{E}_t\bigg|_S~ &=&~ E_{TE_i}\left(1 + \Gamma_{TE}\right)(\mbold{\hat{e}}_{TE}\times\mbold{\hat{n}}) \label{eq:Mmeca} \\
& &+&~ E_{TM_i}\cos(\theta_i)\left(1 + \Gamma_{TM}\right)\mbold{\hat{e}}_{TE} \bigg|_S \nonumber
\end{alignat}
where $\mbold{\hat{n}}$ is the normal of the triangular surface, $\mbold{H}_t$ and $\mbold{E}_t$ are the total magnetic and electric fields, $\eta_1$ is the impedance of the background medium, $E_{TE_i}$ and $E_{TM_i}$ are the TE / TM components of the incident electric field, $\mbold{\hat{e}}_{TE}$ is the unit vector for the TE component of the electric field, $\theta_i$ is the incident angle, and $\Gamma_{TE}$ and $\Gamma_{TM}$ are the TE / TM reflection coefficients. Assuming that the background medium is homogeneous, the scattered electric fields can be computed as follows:
\begin{align}
\mbold{E}_s(\mbold{r}) &= \frac{\jmath}{2\lambda_1 \|\mbold{r}\|_{\ell_2}}e^{-\jmath k_1 \|\mbold{r}\|_{\ell_2}} \label{eq:Escat} \\
& \times \int_S [\mbold{\hat{r}} \times \mbold{M} - (\mbold{\hat{r}} \times \eta_1 \mbold{J} \times \mbold{\hat{r}})]e^{\jmath k_1 \mbold{\hat{r}} \cdot \mbold{r}'} \delta \mbold{r}' \nonumber
\end{align}
where $k_1$ and $\lambda$ are the wavenumber and wavelength of the background medium, $\mbold{r}$ is the point at which the scattered fields are evaluated, $\mbold{\hat{r}} = \frac{\mbold{r}}{\|\mbold{r}\|_{\ell_2}}$, and $\mbold{r}'$ is a vector pointing from the center of the triangular surface to another point on the surface. 

The terms in Eq. \ref{eq:Jmeca} - \ref{eq:Escat} can be separated into two groups: those that vary with the reflection coefficients, and those that do not. Taking this into consideration, we can express the discretized fields $\mbold{g}_{t,i,m} \in \mathbb{C}^N$ radiated by the transmitting antenna as follows:
\begin{align}
\mbold{g}_{t,i,m}(\mvar{x}) = \mbold{e}_{t,i,m}(\mvar{x}) &+ \mbold{E}_{TE,t,i,m}(\mvar{x})\mbold{\Gamma}_{TE,t,m}(\mvar{x}) \\
&+ \mbold{E}_{TM,t,i,m}(\mvar{x})\mbold{\Gamma}_{TM,t,m}(\mvar{x}) \nonumber
\end{align}
where $\mbold{e}_{t,i,m}(\mvar{x}) \in \mathbb{C}^N$, $\mbold{E}_{TE,t,i,m}(\mvar{x})$ and $\mbold{E}_{TM,t,i,m}(\mvar{x}) \in \mathbb{C}^{N \times L}$, $\mbold{\Gamma}_{TE,t,m}(\mvar{x})$ and $\mbold{\Gamma}_{TM,t,m}(\mvar{x}) \in \mathbb{C}^L$, and $L$ is the number of triangular facets on the CRA. Similar expressions can be obtained for the receiving antennas, simply replace the $t$ subscript with $r$. In the most general scenario, $\mbold{e}_{t,i,m}(\mvar{x})$, $\mbold{E}_{TE,t,i,m}(\mvar{x})$, $\mbold{E}_{TM,t,i,m}(\mvar{x})$, $\mbold{\Gamma}_{TE,t,m}(\mvar{x})$, and $\mbold{\Gamma}_{TM,t,m}(\mvar{x})$ can all vary as a function of the design variables $\mvar{x}$. Assuming that they are all differentiable, the partial derivatives $\frac{\partial \mbold{g}_{t,i,m}(\mvar{x})}{\partial x_p}$ can be expressed as follows: 
\begin{align}
\frac{\partial \mbold{g}_{t,i,m}(\mvar{x})}{\partial x_p} = \frac{\partial \mbold{e}_{t,i,m}(\mvar{x}) }{\partial x_p} &+ \frac{\partial \mbold{E}_{TE,t,i,m}(\mvar{x}) }{\partial x_p}\mbold{\Gamma}_{TE,t,m}(\mvar{x}) \nonumber \\
&+  \mbold{E}_{TE,t,i,m}(\mvar{x})\frac{\partial \mbold{\Gamma}_{TE,t,m}(\mvar{x}) }{\partial x_p} \nonumber \\
&+ \frac{\partial \mbold{E}_{TM,t,i,m}(\mvar{x}) }{\partial x_p}\mbold{\Gamma}_{TM,t,m}(\mvar{x}) \nonumber \\
&+  \mbold{E}_{TM,t,i,m}(\mvar{x})\frac{\partial \mbold{\Gamma}_{TM,t,m}(\mvar{x}) }{\partial x_p} 
\end{align}

\subsection{Dielectric CRA Formulation}
Consider the scenario in which the design variables are the dielectric constant of purely dielectric scatterers. In this case, the fields $\mbold{g}_{t,i,m}(\mvar{x})$ and their derivatives can be expressed as follows:
\begin{align}
\mbold{g}_{t,i,m}(\mvar{x}) = \mbold{e}_{t,i,m} &+ \mbold{E}_{TE,t,i,m}\mbold{\Gamma}_{TE,t,m}(\mvar{x})  \\
&+ \mbold{E}_{TM,t,i,m}\mbold{\Gamma}_{TM,t,m}(\mvar{x}) \nonumber \\
\frac{\partial \mbold{g}_{t,i,m}(\mvar{x})}{\partial x_p} = &~\mbold{E}_{TE,t,i,m}\frac{\partial \mbold{\Gamma}_{TE,t,m}(\mvar{x}) }{\partial x_p} + \\
&~\mbold{E}_{TM,t,i,m}\frac{\partial \mbold{\Gamma}_{TM,t,m}(\mvar{x}) }{\partial x_p} \nonumber
\end{align}
Note that the Jacobian matrices $\frac{\partial \mbold{\Gamma}_{TE,t,m}(\mvar{x}) }{\partial \mvar{x}}$ and $\frac{\partial \mbold{\Gamma}_{TM,t,m}(\mvar{x}) }{\partial \mvar{x}}$ are diagonal because the number of reflection coefficients $L$ equals the number of design variables $P$. The reflection coefficients are computed by treating each facet of the CRA as a three-layer medium containing the background medium (i.e. freespace), the dielectric scatterer, and the perfect electric conductor (PEC) reflector. MECA considers the incident fields to be locally planar at each facet, such that the reflection coefficient of the $p-$th scatterer can be expressed as follows:
\begin{align}
\Gamma_{m,p}(\epsilon_p) &= \Gamma_{bp,m}(\epsilon_p)  \label{eq:gamma_tlm}\\
&-\frac{T_{bp,m}(\epsilon_p)T_{pb,m}(\epsilon_p)e^{-\jmath 2 \phi_{m,p}(\epsilon_p)}}{1 + \Gamma_{pb,m}(\epsilon_p)e^{-\jmath 2 \phi_{m,p}(\epsilon_p)}} \nonumber \\
\phi_{m,p}(\epsilon_p) &= 2\pi f_m \sqrt{\mu_0\epsilon_0\epsilon_p}d_p\cos(\theta_{t_t,m,p}(\epsilon_p)) \label{eq:phase_delay}
\end{align}
where $\Gamma_{bp,m}(\epsilon_p)$ and $T_{bp,m}(\epsilon_p)$ are the half-plane reflection and transmission coefficients going from the background medium to the dielectric, $\Gamma_{pb,m}(\epsilon_p)$ and $T_{pb,m}(\epsilon_p)$ are the half-plane reflection and transmission coefficients going from the dielectric to the background medium, $f_m$ is the frequency, $d_p$ is the thickness of the dielectric, and $\theta_{t_t,m,p}(\epsilon_p)$ is the angle of propagation in the dielectric. Note that we have replaced the generic design variable $x_p$ with the dielectric constant $\epsilon_p$ for clarity. Eq. \ref{eq:gamma_tlm} is a simplified version of the general relationship described in \cite{Chew1995} for the case where the third medium is a PEC. It is straightforward to show that the derivatives of Eq. \ref{eq:gamma_tlm} and \ref{eq:phase_delay} can be expressed as follows:
\begin{align}
\frac{d\Gamma_{m,p}(\epsilon_p)}{d \epsilon_p} &= \frac{d\Gamma_{bp,m}(\epsilon_p)}{d \epsilon_p} \label{eq:d_gamma_tlm} \\
&-\left(1 + \Gamma_{pb,m}(\epsilon_p)e^{-\jmath 2 \phi_{m,p}(\epsilon_p)}\right)^{-1} \nonumber \\
&\times\bigg(\frac{dT_{bp,m}(\epsilon_p)}{d \epsilon_p}T_{pb,m}(\epsilon_p)e^{-\jmath 2 \phi_{m,p}(\epsilon_p)} \nonumber \\
&+T_{bp,m}(\epsilon_p)\frac{d T_{pb,m}(\epsilon_p)}{d \epsilon_p}e^{-\jmath 2 \phi_{m,p}(\epsilon_p)} \nonumber \\
&-\jmath 2 T_{bp,m}(\epsilon_p)T_{pb,m}(\epsilon_p)\frac{d \phi_{m,p}(\epsilon_p)}{d \epsilon_p}e^{-\jmath 2 \phi_{m,p}(\epsilon_p)}\bigg) \nonumber \\
&+\left(T_{bp,m}(\epsilon_p)T_{pb,m}(\epsilon_p)e^{-\jmath 2 \phi_{m,p}(\epsilon_p)}\right) \nonumber \\
&\times\left(1 + \Gamma_{pb,m}(\epsilon_p)e^{-\jmath 2 \phi_{m,p}(\epsilon_p)}\right)^{-2} \nonumber \\
&\times\bigg( \frac{d\Gamma_{pb,m}(\epsilon_p)}{d \epsilon_p} e^{-\jmath 2 \phi_{m,p}(\epsilon_p)} \nonumber \\
&-\jmath 2 \Gamma_{pb,m}(\epsilon_p)\frac{d \phi_{m,p}(\epsilon_p)}{d \epsilon_p}e^{-\jmath 2 \phi_{m,p}(\epsilon_p)}\bigg) \nonumber
\end{align}
\begin{align}
\frac{d\phi_{m,p}(\epsilon_p) }{d \epsilon_p} &= \pi f_m \sqrt{\mu_0\epsilon_0}d_p\bigg(\frac{1}{\sqrt{\epsilon_p}}\cos(\theta_{t_t,m,p}(\epsilon_p)) \label{eq:d_phase_delay} \\
&+ 2\sqrt{\epsilon_p}\frac{d \cos(\theta_{t_t,m,p}(\epsilon_p) )}{d\epsilon_p}\bigg)  \nonumber
\end{align}

To evaluate these equations, we require expressions for the transmission coefficients, reflection coefficients, and the propagation angle $\theta_{t_t,m,p}(\epsilon_p)$. The transmission and reflection coefficient formulations differ for the TE and TM modes, so we first consider the propagation angle. The following relationships for $\cos(\theta_{t_t,m,p}(\epsilon_p))$ are easily found by invoking Snell's Law:
\begin{align}
\cos(\theta_{t_t,m,p}(\epsilon_p)) &= \sqrt{1 - \frac{\epsilon_b}{\epsilon_p}\sin^2(\theta_{i_t,m,p})} \\
\frac{d \cos(\theta_{t_t,m,p}(\epsilon_p)) }{d \epsilon_p} &=\frac{\epsilon_b\sin^2(\theta_{i_t,m,p})}{2 \epsilon_p^2 \sqrt{1-\frac{\epsilon_b}{\epsilon_p}\sin^2(\theta_{i_t,m,p})}}
 \end{align} 
where $\epsilon_b$ is the dielectric constant of the homogeneous background medium and $\theta_{i_t,m,p}$ is the incident angle. 

To simplify the notation when evaluating the transmission and reflection coefficients, we will drop the subscripts $m$ (measurement index), $t$ (transmitting antenna), and $p$ (dielectric scatterer index) on some of the parameters with the understanding that the formulations can be applied to all measurements and scatterers independently. According to electromagnetic theory, the half-plane TM reflection and transmission coefficients can be expressed as follows \cite{ulaby2010fundamentals}:
\begin{align}
\Gamma_{bp,TM} &= \frac{\cos(\theta_i) - \sqrt{\frac{\epsilon_p}{\epsilon_b}-\sin^2(\theta_i)}}{\cos(\theta_i) + \sqrt{\frac{\epsilon_p}{\epsilon_b}-\sin^2(\theta_i)}} \label{eq:ref_bp_tm} \\
T_{bp,TM} &= 1 + \Gamma_{bp,TM}  \label{eq:trans_bp_tm}
\end{align}
It's easy to show from Eq. \ref{eq:ref_bp_tm} that the derivative $\frac{d \Gamma_{bp,TM}}{d\epsilon_p}$ can be expressed as follows:
\begin{align}
\frac{d \Gamma_{bp,TM}}{d\epsilon_p} &= \frac{-1}{2\epsilon_b\sqrt{\frac{\epsilon_p}{\epsilon_b}-\sin^2(\theta_i)}\left(\cos(\theta_i)+ \sqrt{\frac{\epsilon_p}{\epsilon_b}-\sin^2(\theta_i)}\right)}\nonumber  \\
&-\frac{\cos(\theta_i)+ \sqrt{\frac{\epsilon_p}{\epsilon_b}-\sin^2(\theta_i)}}{2\epsilon_b \left(\cos(\theta_i)+ \sqrt{\frac{\epsilon_p}{\epsilon_b}-\sin^2(\theta_i)}\right)^2 \sqrt{\frac{\epsilon_p}{\epsilon_b}-\sin^2(\theta_i)}}  \label{eq:d_ref_bp_tm}
\end{align}
Clearly, $\frac{d T_{bp,TM}}{d\epsilon_p} = \frac{d \Gamma_{bp,TM}}{d\epsilon_p}$, so it is omitted. The coefficients $\Gamma_{pb,TM}$ and $T_{pb,TM}$ and their derivatives can easily be found by exploiting the fact that $\Gamma_{pb,TM} = -\Gamma_{bp,TM}$, so they are also omitted.

The TE reflection and transmission coefficients can be expressed as follows \cite{ulaby2010fundamentals}:
\begin{align}
\Gamma_{bp,TE} &= \frac{-\frac{\epsilon_p}{\epsilon_b}\cos(\theta_i) + \sqrt{\frac{\epsilon_p}{\epsilon_b}-\sin^2(\theta_i)}}{\frac{\epsilon_p}{\epsilon_b}\cos(\theta_i) + \sqrt{\frac{\epsilon_p}{\epsilon_b}-\sin^2(\theta_i)}} \label{eq:ref_bp_te} \\
T_{bp,TE} &= \left(1 + \Gamma_{bp,TE}\right)\frac{\cos(\theta_i)}{\cos(\theta_t)}  \label{eq:trans_bp_te} \\
&=  \left(1 + \Gamma_{bp,TE}\right)\frac{\cos(\theta_i)}{\sqrt{1-\frac{\epsilon_b}{\epsilon_p}\sin^2(\theta_i)}}  \nonumber
\end{align}
We first turn our attention to the reflection coefficient. The derivative $\frac{d \Gamma_{bp,TE}}{d\epsilon_p}$ can be expressed as follows:
\begin{align}
\frac{d \Gamma_{bp,TE}}{d\epsilon_p} &= \frac{-\cos(\theta_i) + \frac{1}{2\sqrt{\frac{\epsilon_p}{\epsilon_b}-\sin^2(\theta_i)}}}{\epsilon_p\cos(\theta_i) + \epsilon_b\sqrt{\frac{\epsilon_p}{\epsilon_b}-\sin^2(\theta_i)}} \label{eq:d_ref_bp_te} \\
&-\left(-\epsilon_p\cos(\theta_i) + \epsilon_b\sqrt{\frac{\epsilon_p}{\epsilon_b}-\sin^2(\theta_i)}\right)\nonumber \\
&\times\left(\frac{1}{2\sqrt{\frac{\epsilon_p}{\epsilon_b} - \sin^2(\theta_i)}} + \cos(\theta_i)\right) \nonumber \\
&\times\left(\epsilon_p\cos(\theta_i) + \epsilon_b\sqrt{\frac{\epsilon_p}{\epsilon_b}-\sin^2(\theta_i)}\right)^{-2} \nonumber
\end{align}
Just as in the TM case, the reflection coefficient $\Gamma_{pb,TE}$ and its derivative can be obtained using the relationship $\Gamma_{pb,TE}=-\Gamma_{bp,TE}$. 

We now turn our attention to the transmission coefficients. The derivative $\frac{d T_{bp,TE}}{d\epsilon_p}$ can be expressed as follows:
\begin{align}
\frac{d T_{bp,TE}}{d\epsilon_p} &= \frac{\cos(\theta_i)\frac{d \Gamma_{bp,TE}}{d\epsilon_p}}{\sqrt{1-\frac{\epsilon_b}{\epsilon_p}\sin^2(\theta_i)}}  \label{eq:d_trans_bp_te} \\
&- \frac{\epsilon_b\cos(\theta_i)\sin^2(\theta_i)(1+\Gamma_{bp,TE})}{2\epsilon_p^2\left(1-\frac{\epsilon_b}{\epsilon_p}\sin^2(\theta_i)\right)^{\frac{3}{2}}}\nonumber
\end{align}
To compute $\frac{d T_{pb,TE}}{d\epsilon_p}$, we cannot simply replace the $\Gamma_{bp,TE}$ terms with $\Gamma_{pb,TE}$ due to the presence of the $\theta_t$ term in Eq. \ref{eq:ref_bp_te}. This result is clear from Eq. \ref{eq:trans_pb_te}:
\begin{align}
T_{pb,TE} &= \left(1 + \Gamma_{pb,TE}\right)\frac{\cos(\theta_t)}{\cos(\theta_e)}  \label{eq:trans_pb_te} \\
&= \left(1 + \Gamma_{pb,TE}\right)\frac{\sqrt{1-\frac{\epsilon_b}{\epsilon_p}\sin^2(\theta_i)}}{\cos(\theta_i)} \nonumber
\end{align}
It is now easy to show the following expression for $\frac{d T_{pb,TE}}{d\epsilon_p}$:
\begin{align}
\frac{d T_{pb,TE}}{d\epsilon_p} &= \frac{-\frac{d \Gamma_{bp,TE}}{d\epsilon_p}\sqrt{1-\frac{\epsilon_b}{\epsilon_p}\sin^2(\theta_i)}}{\cos(\theta_i)} \label{eq:d_trans_pb_te} \\
&+ \frac{\left(1-\Gamma_{bp,TE}\right)\epsilon_b\sin^2(\theta_i)}{2\epsilon_p^2\cos(\theta_i)\sqrt{1 - \frac{\epsilon_b}{\epsilon_p}\sin^2(\theta_i)}} \nonumber
 \end{align} 




\vspace{7pt}
\section{Design Results}
\label{sec:results}
\subsection{Compressive Imaging Application}
\begin{figure}[h!]
\centering
\includegraphics[width=7.5cm, clip=true]{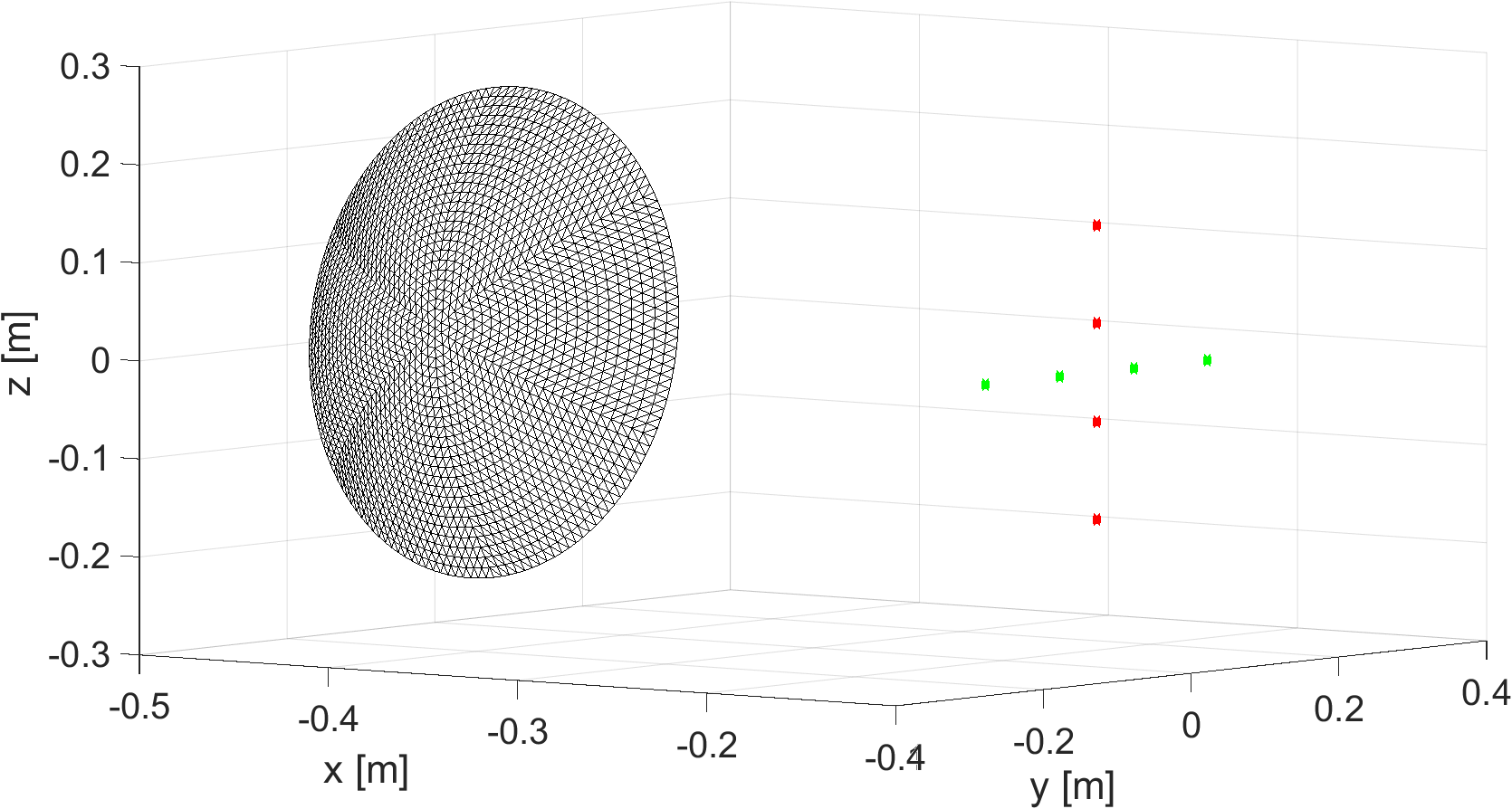}
    \caption{Test configuration for the design problem. Four transmitting (green dots) and four receiving antennas (red dots) were used in conjunction with the dielectric CRA reflector located at $x=-0.5$[m].}
   \label{fig:design_geom}
\end{figure}

The dielectric CRA design method has been assessed using the test configuration displayed in Figure \ref{fig:design_geom}. In the design scenario, four receiving antennas (red dots) were used to measure the signals generated by the four transmitting antennas (green dots). The antennas operated in a multi-static configuration at ten frequencies uniformly spaced between $70.5$[GHz] and $77$[GHz], for a total of $160$ measurements. The antennas were modeled as horn antennas, wherein the incident electric field was oriented in the $+z$ direction and the direction of propagation was the $-x$ direction. The parabolic CRA reflector, which was located at $x=-0.5$[m], was discretized into $3750$ triangular facets. The design procedure was configured to optimize the dielectric constant of each triangular facet, wherein $\epsilon_r \in [1, 30]$. All dielectrics had the same thickness $d=8.13$[mm], which is approximately two times the average wavelength of the operating frequencies. The imaging region consisted of $625$ positions uniformly spaced along a $0.2$[m] $\times$ $0.2$[m] grid centered at $x=0.9$[m]. The optimization problem of Eq. \ref{eq:unified_opt} was executed four times using the parameters listed in Table \ref{tab:tab1}. The first three configurations correspond to pure-capacity, pure-efficiency, and joint capacity and efficiency optimization of the sensing matrix, and the final configuration optimizes over all terms. To initialize the optimization procedure, the starting dielectric constant values were randomly selected over the feasible set.

\begin{table}[!b]
\captionsetup{font=normal}
\centering
\begin{tabular}{ |C{1.3cm}||C{1.3cm}||C{1.3cm}||C{1.3cm}||C{1.3cm}|  }
 \hline
 \multicolumn{5}{|c|}{Design Parameters} \\
 \hline
& Capacity Opt. & Efficiency Opt. & Capacity and Efficiency Opt. & All Terms Opt. \\
 \hline
$\alpha_A$ & $1$ & $0$ & $1$ & $1$ \\
\hline
$\alpha_{r_i}$ & $0$ & $0$ & $0$ & $0.25$ \\
\hline
$\alpha_{t_i}$ & $0$ & $0$ & $0$ & $0.25$ \\
\hline
$\beta_A$ & $10^{-6}$ & $10^{-6}$ & $10^{-6}$ & $10^{-6}$ \\
\hline
$\beta_{r_i}$ & $10^{-6}$ & $10^{-6}$ & $10^{-6}$ & $10^{-6}$ \\
\hline
$\beta_{t_i}$ & $10^{-6}$ & $10^{-6}$ & $10^{-6}$ & $10^{-6}$ \\
\hline
$\mbold{\Lambda}_{A,m}$ & $\mbold{0}_N$ & $\mbold{I}_N$ & $10\mbold{I}_N$ & $10\mbold{I}_N$ \\
\hline
$\mbold{\Lambda}_{r,i,m}$ & $\mbold{0}_N$ & $\mbold{0}_N$ & $\mbold{0}_N$ & $2.5\mbold{I}_N$ \\
\hline
$\mbold{\Lambda}_{ti,m}$ & $\mbold{0}_N$ & $\mbold{0}_N$ & $\mbold{0}_N$ & $2.5\mbold{I}_N$ \\
 \hline
\end{tabular}
\caption{Optimization parameters for the designs.}
\label{tab:tab1}
\end{table}

\begin{figure}[h!]
\centering
\includegraphics[width=7.5cm, clip=true]{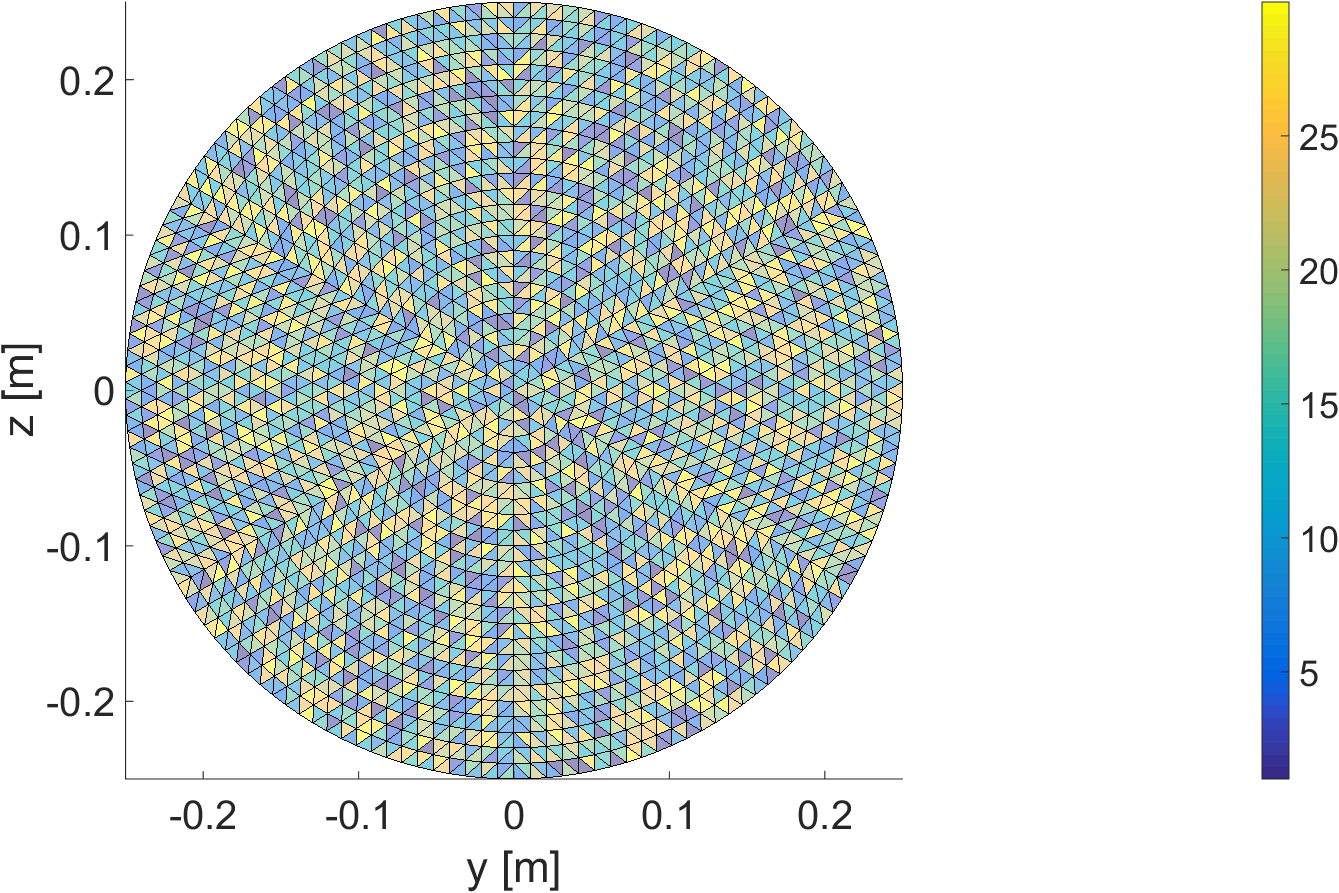}
    \caption{Dielectric constant of each triangular facet on the baseline randomized CRA.}
   \label{fig:random_reflector}
\end{figure}

\begin{figure}[h!]
\centering
\includegraphics[width=7.5cm, clip=true]{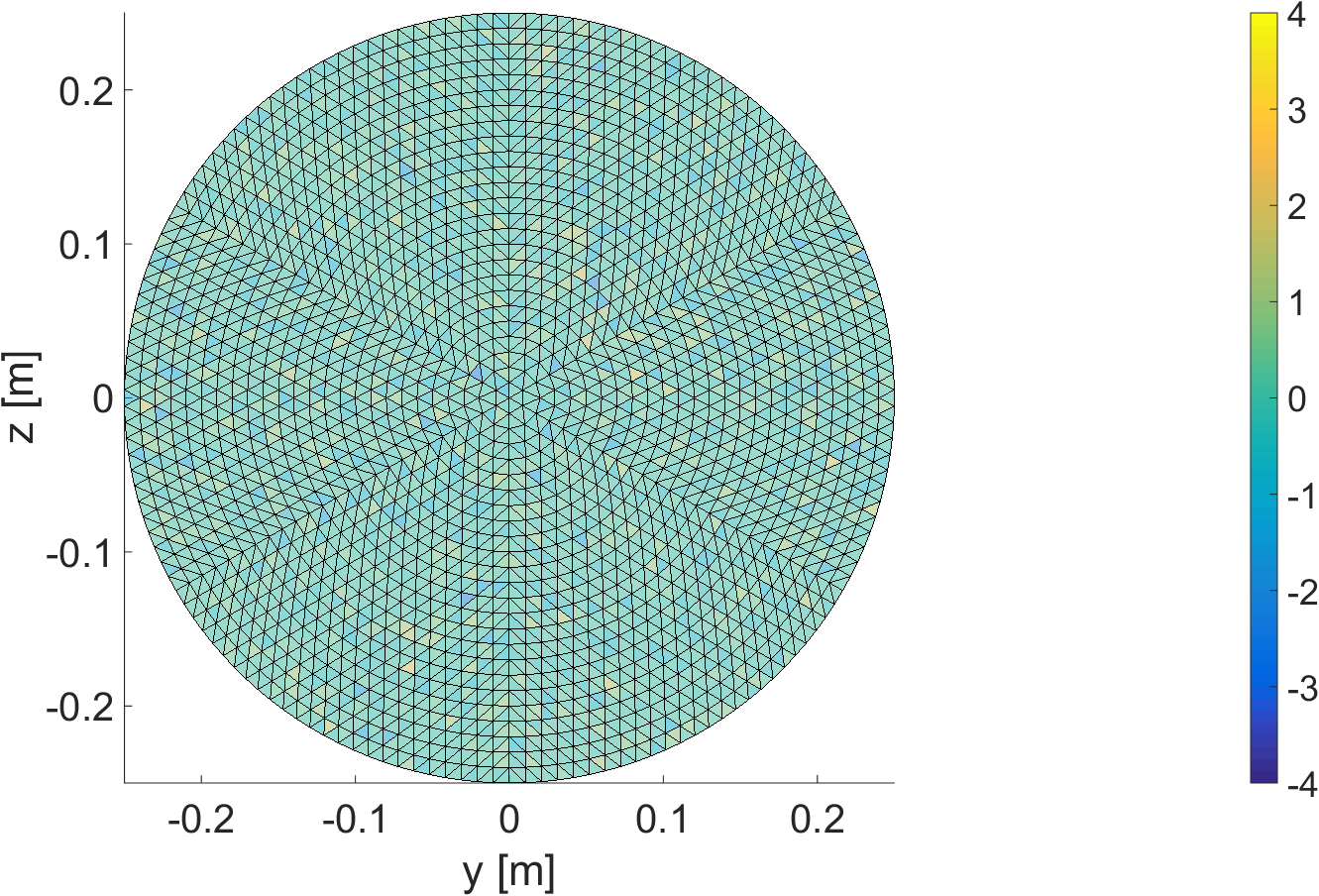}
    \caption{Difference in dielectric constant between the optimized capacity CRA and the baseline randomized CRA.}
   \label{fig:capacity_difference_result}
\end{figure}

\begin{figure}[h!]
\centering
\includegraphics[width=7.5cm, clip=true]{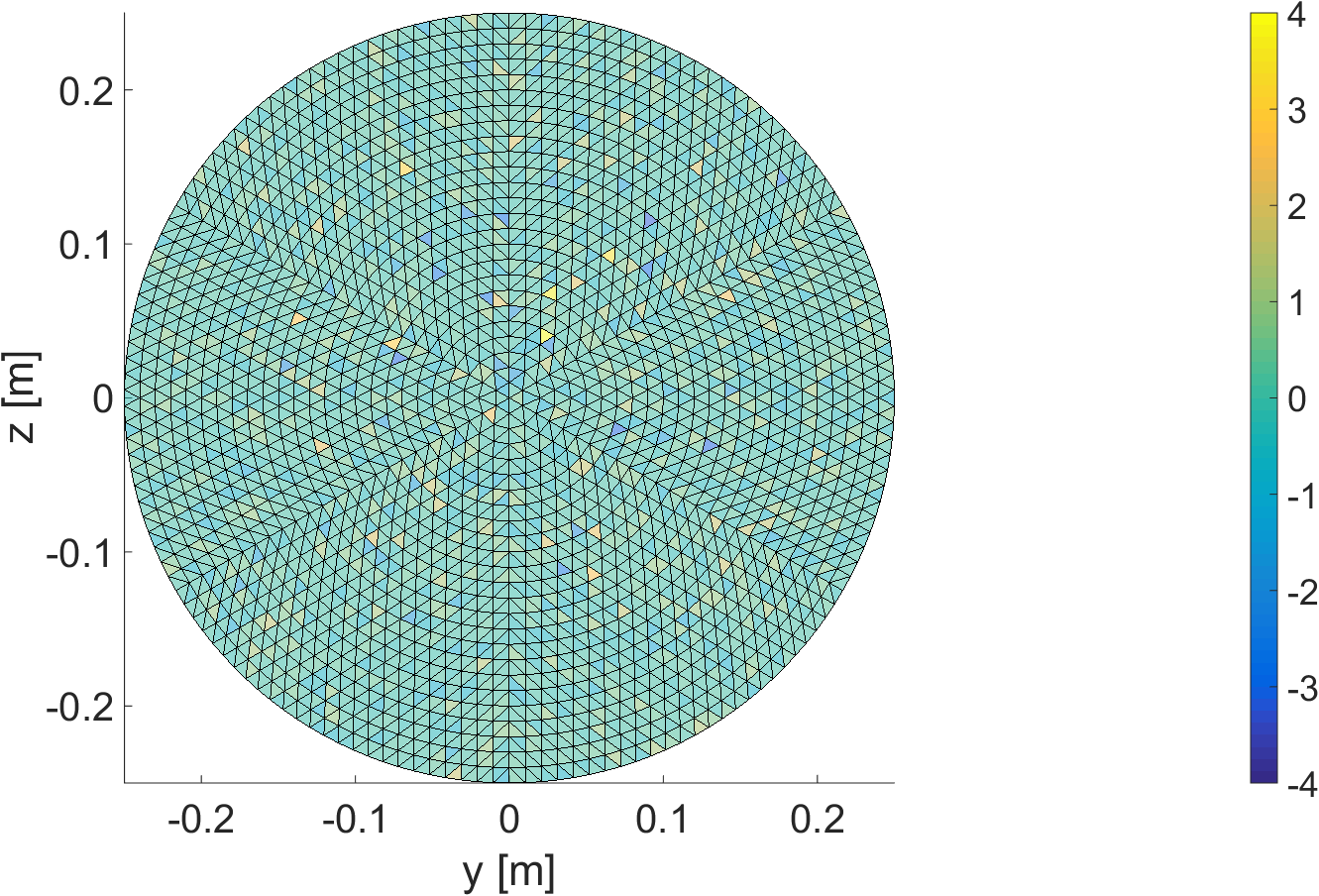}
    \caption{Difference in dielectric constant between the optimized efficiency CRA and the baseline randomized CRA.}
   \label{fig:efficiency_difference_result}
\end{figure}

\begin{figure}[h!]
\centering
\includegraphics[width=7.5cm, clip=true]{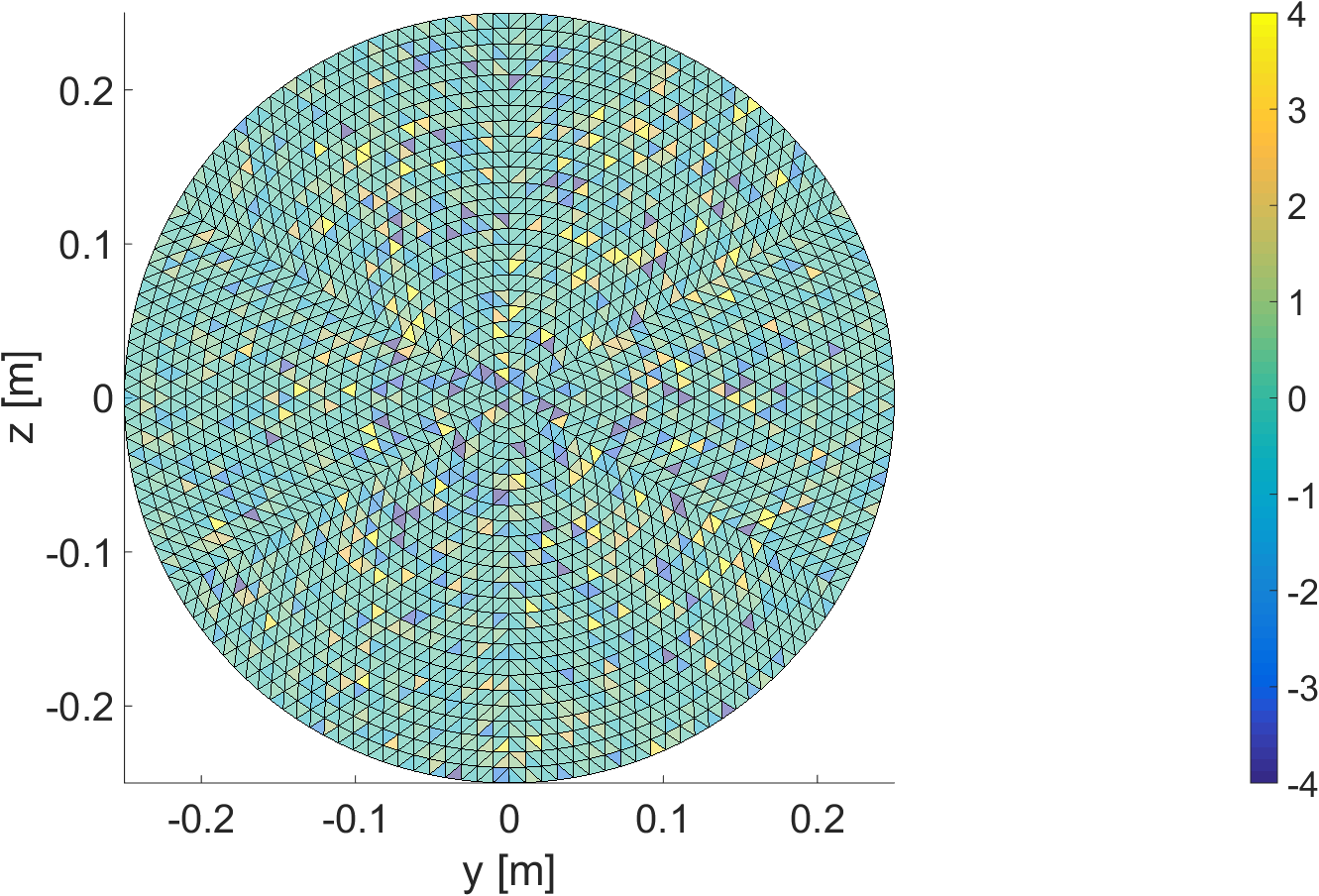}
    \caption{Difference in dielectric constant between the optimized capacity plus efficiency CRA and the baseline randomized CRA.}
   \label{fig:capacity_plus_efficiency_difference_result}
\end{figure}

\begin{figure}[h!]
\centering
\includegraphics[width=7.5cm, clip=true]{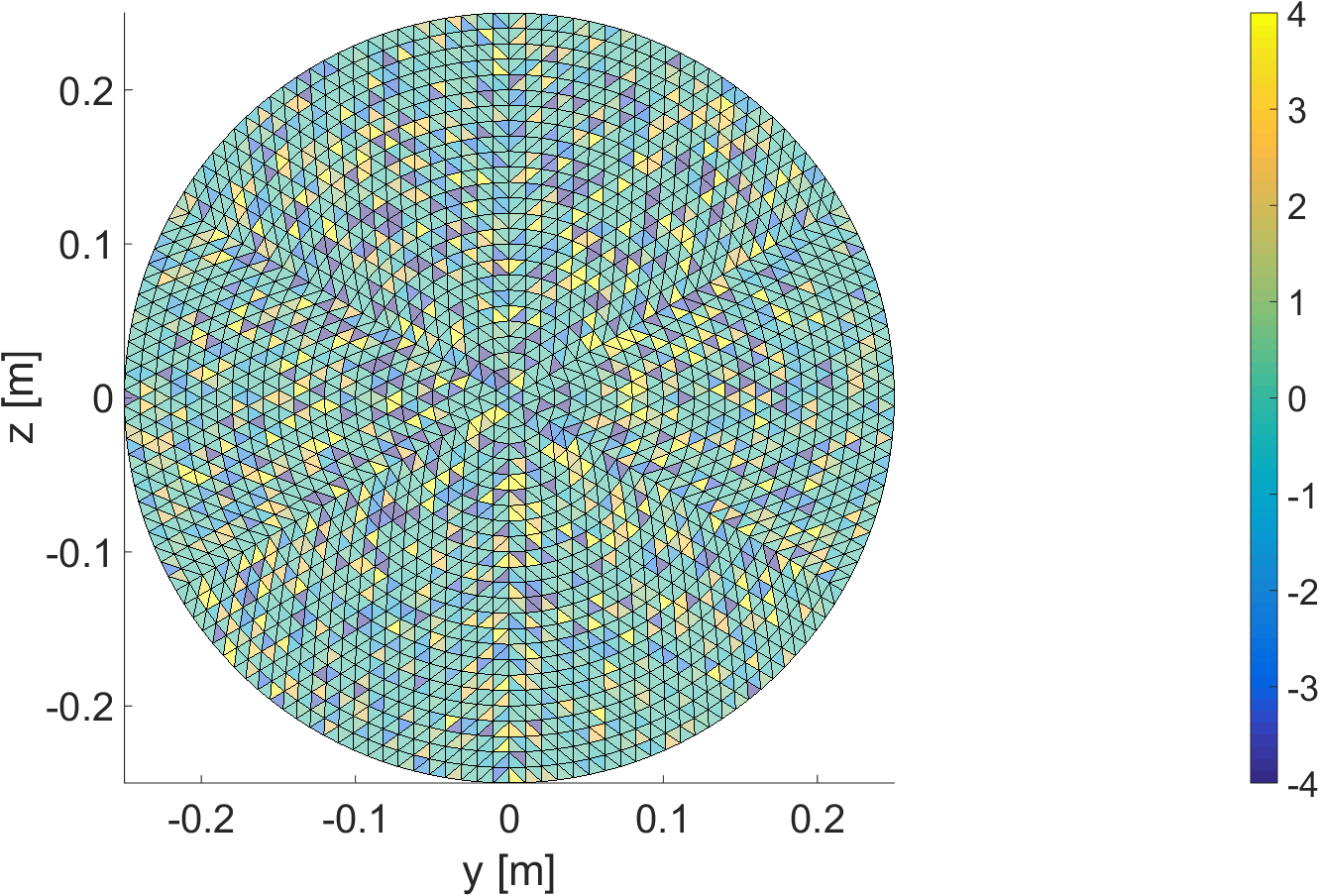}
    \caption{Difference in dielectric constant between the CRA optimized using all terms and the baseline randomized CRA.}
   \label{fig:capacity_all_terms_difference_result}
\end{figure}

\begin{table}[h!]
\captionsetup{font=normal}
\centering
\begin{tabular}{ |C{2.9cm}||C{1.3cm}||C{1.3cm}||C{1.3cm}| }
 \hline
 \multicolumn{4}{|c|}{Capacity} \\
 \hline
& $\mbold{A}$ & $\mbold{G}_r$ & $\mbold{G}_t$ \\
\hline
\multirow{2}{*}{TRA}  & -1786.4  & -6038.7 & -5646.7  \\
\cline{2-4}
  &  \textcolor{black}{-713.2} & \textcolor{black}{-399.8} & \textcolor{black}{-494.8} \\
 \hline 
\multirow{2}{*}{Random CRA}  & -1410.8  & -5747.1  & -5297.3  \\
\cline{2-4}
&  \textcolor{black}{-337.6} &  \textcolor{black}{-108.3} & \textcolor{black}{-145.4} \\
 \hline 
\multirow{2}{*}{Capacity Opt.}  & -1098.2  & -5652.4  & -5183.3 \\
\cline{2-4}
&  \textcolor{black}{-25.0} & \textcolor{black}{-13.5} &\textcolor{black}{-31.5} \\
 \hline 
\multirow{2}{*}{Efficiency Opt.}  & -1179.7  & -5699.6 & -5223.0 \\
\cline{2-4}
& \textcolor{black}{-106.5} &  \textcolor{black}{-60.7} &  \textcolor{black}{-71.2} \\
 \hline 
\multirow{2}{*}{Capacity / Efficiency Opt.}  & -1121.5  & -5674.8  & -5201.8  \\
\cline{2-4}
&  \textcolor{black}{-48.3} & \textcolor{black}{-35.9} &  \textcolor{black}{-49.9} \\
 \hline 
\multirow{2}{*}{All Terms Opt.}  & -1073.2  & -5638.9  & -5151.9  \\
\cline{2-4}
& \textcolor{black}{0.0} &  \textcolor{black}{0.0} & \textcolor{black}{0.0} \\
 \hline 
\end{tabular}
\caption{Capacity of the sensing matrix and Green's functions for each design scenario. For the Green's functions, all three components are added together. The top number in each row is the achieved value, and the bottom number is the difference between that value and the best achieved value.}
\label{tab:cap_sum}
\end{table}

\begin{table}[h!]
\captionsetup{font=normal}
\centering
\begin{tabular}{ |C{2.9cm}||C{1.3cm}||C{1.3cm}||C{1.3cm}| }
 \hline
 \multicolumn{4}{|c|}{Efficiency} \\
 \hline
& $\mbold{A}$ & $\mbold{G}_r$ & $\mbold{G}_t$ \\
\hline
\multirow{2}{*}{TRA}  & 0.3  & 169.9  & 88.4   \\
\cline{2-4}
  &  \textcolor{black}{-522.6} &  \textcolor{black}{-2295.6} &  \textcolor{black}{-1784.8} \\
 \hline 
\multirow{2}{*}{Random CRA}  & 4.2  & 629.4  & 499.5  \\
\cline{2-4}
&  \textcolor{black}{-518.8} & \textcolor{black}{-1836.1} & \textcolor{black}{-1373.6} \\
 \hline 
\multirow{2}{*}{Capacity Opt.}  & 53.8  & 1939.8 & 1496.9  \\
\cline{2-4}
& \textcolor{black}{-469.2} & \textcolor{black}{-525.7} & \textcolor{black}{-376.3} \\
 \hline 
\multirow{2}{*}{Efficiency Opt.}  & 522.9 & 1548.5  & 1185.2 \\
\cline{2-4}
& \textcolor{black}{0.0} &  \textcolor{black}{-917.0} & \textcolor{black}{-688.0}\\
 \hline 
\multirow{2}{*}{Capacity / Efficiency Opt.} & 287.6  & 1745.9  & 1331.8  \\
\cline{2-4}
&  \textcolor{black}{-235.3} &  \textcolor{black}{-719.5} &  \textcolor{black}{-541.4}\\
 \hline 
\multirow{2}{*}{All Terms Opt.}  & 70.6 & 2465.5 & 1873.1  \\
\cline{2-4}
& \textcolor{black}{-452.3} &  \textcolor{black}{0.0} &  \textcolor{black}{0.0} \\
 \hline 
\end{tabular}
\caption{Efficiency of the sensing matrix and Green's functions for each design scenario. For the Green's functions, all three components are added together. The top number in each row is the achieved value, and the bottom number is the difference between that value and the best achieved value.}
\label{tab:eff_sum}
\end{table}


Figure \ref{fig:random_reflector} displays the dielectric constant of each triangular facet on the baseline randomized CRA, and Figures \ref{fig:capacity_difference_result} - \ref{fig:capacity_all_terms_difference_result} display the difference in dielectric constant between the optimized CRA's and the baseline randomized CRA. The change in dielectric constant varied for each of the design configurations, with the pure capacity optimization changing the least and the all terms configuration changing the most. Nevertheless, the design algorithm significantly improved the value of the objective function in each case. This can be seen in Tables \ref{tab:cap_sum} and \ref{tab:eff_sum}, which display the capacity and efficiency, respectively, of the sensing matrix and Green's functions for the four design configurations and the baseline randomized CRA as a reference. A TRA is also included for reference. Each table element contains two values, the value of the metric (capacity or efficiency), and the difference between the achieved metric and the best metric achieved by any of the configurations. The optimal sensing matrix efficiency was achieved by optimizing only that metric. The optimal values for all other metrics were achieved by optimizing over all of the terms in Eq. \ref{eq:unified_opt}.

\begin{figure}[h!]
\centering
\includegraphics[width=7.5cm, clip=true]{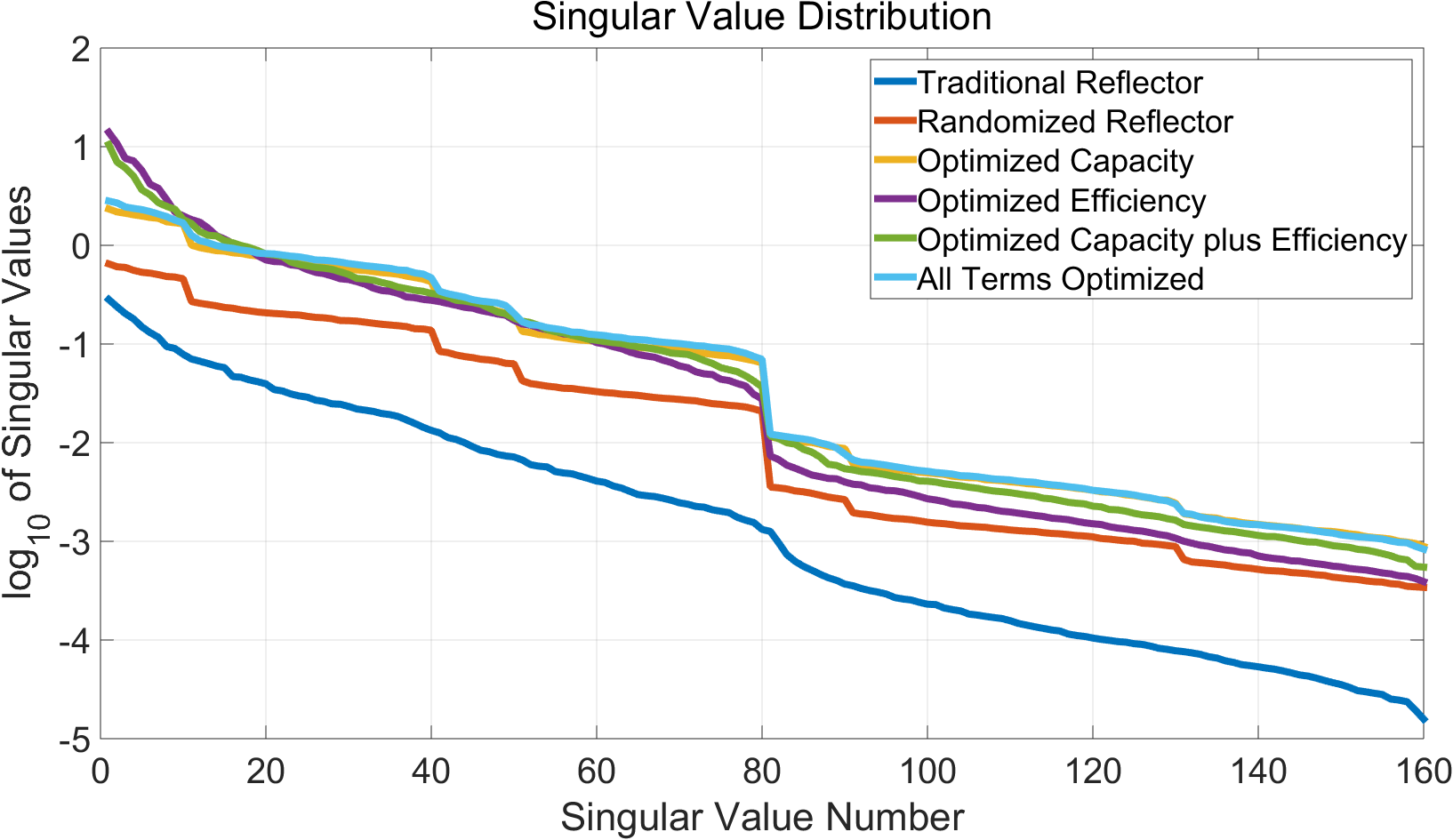}
    \caption{Singular values of the sensing matrices obtained using the TRA, baseline randomized CRA, and optimized CRA's.}
   \label{fig:singular_values}
\end{figure}

Figure \ref{fig:singular_values} displays the singular value distributions of the four optimized designs, the TRA, and the baseline randomized CRA. The TRA unsurprisingly has the worst sensing capacity of the six designs. The four optimized CRA's have improved singular values compared to the randomized starting point, which in turn is an improvement over the TRA. The capacity-optimized and all terms optimized CRA's have singular value distributions that are similar to the randomized CRA, except each singular value is approximately increased by a fixed amount. The efficiency-optimized CRA has the highest maximum singular value, but a less-uniform distribution than the capacity-optimized CRA. The joint capacity and efficiency-optimized CRA falls in between the others.

\begin{figure}[h!]
\centering
\includegraphics[width=7.5cm, clip=true]{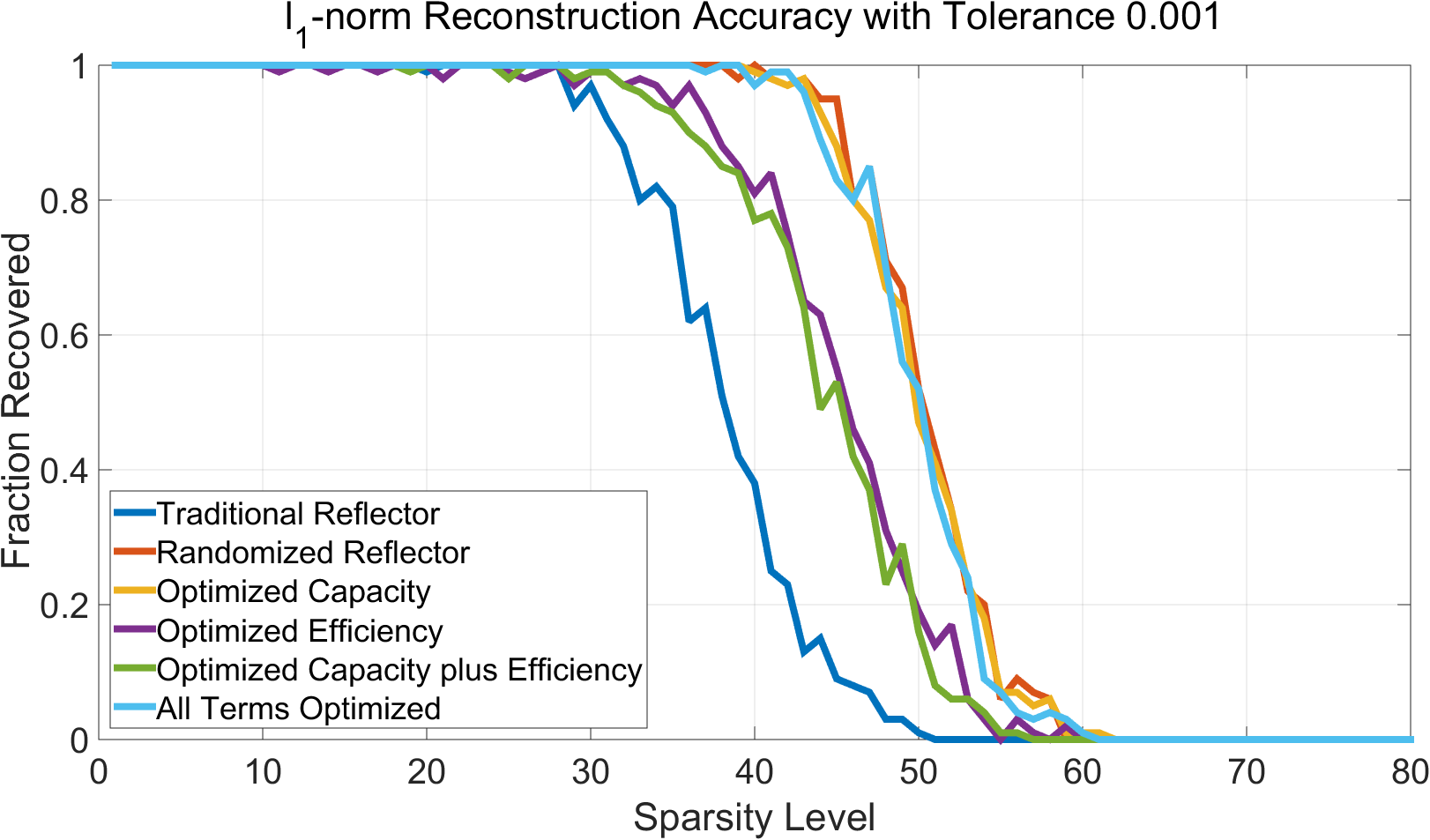}
    \caption{Numerical comparison of the reconstruction accuracies of $\ell_1$-norm minimization using the sensing matrices obtained using the TRA, baseline randomized CRA, and optimized CRA's.}
   \label{fig:norm_one_results}
\end{figure}

\begin{figure}[h!]
\centering
\includegraphics[width=7.5cm, clip=true]{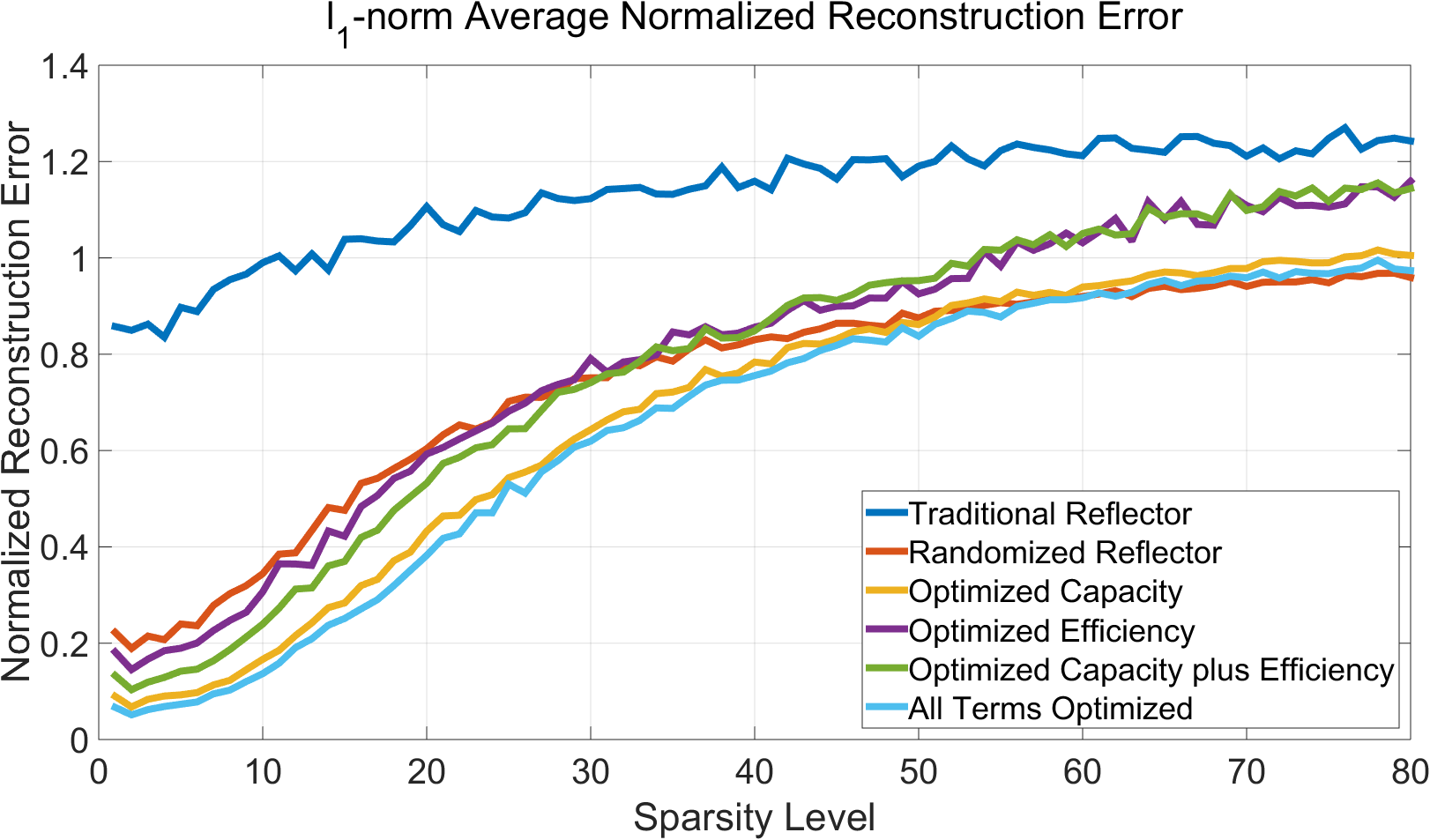}
    \caption{Numerical comparison of the reconstruction accuracies of $\ell_1$-norm minimization using the sensing matrices obtained using the TRA, baseline randomized CRA, and optimized CRA's.}
   \label{fig:norm_one_noisy_results}
\end{figure}

The singular value distributions suggest the following: $1)$ the CRA's demonstrate better CS reconstruction capabilities than the TRA; $2)$ the randomized, capacity-optimized, and all terms optimized CRA's demonstrate similar CS reconstruction capabilities in the absence of measurement noise; $3)$ the capacity-optimized and all terms optimized CRA's demonstrate better CS reconstruction capabilities than the randomized CRA when the measurements are corrupted by noise due to their enhanced capacity. The first and second items are confirmed for this design scenario by Figure \ref{fig:norm_one_results}, which display the $\ell_1$-norm reconstruction capabilities of all antennas in the noiseless scenario. The third item is confirmed by Figure \ref{fig:norm_one_noisy_results}, which displays the $\ell_1$-norm reconstruction capabilities of all antennas in the presence of noise. Unsurprisingly, the efficiency-optimized and the joint capacity and efficiency-optimized CRA's provide a CS reconstruction performance in between the randomized, capacity-optimized, and all terms optimized CRA's. 

\subsection{Communication Application}
\begin{figure}[h!]
\centering
\includegraphics[width=7.5cm, clip=true]{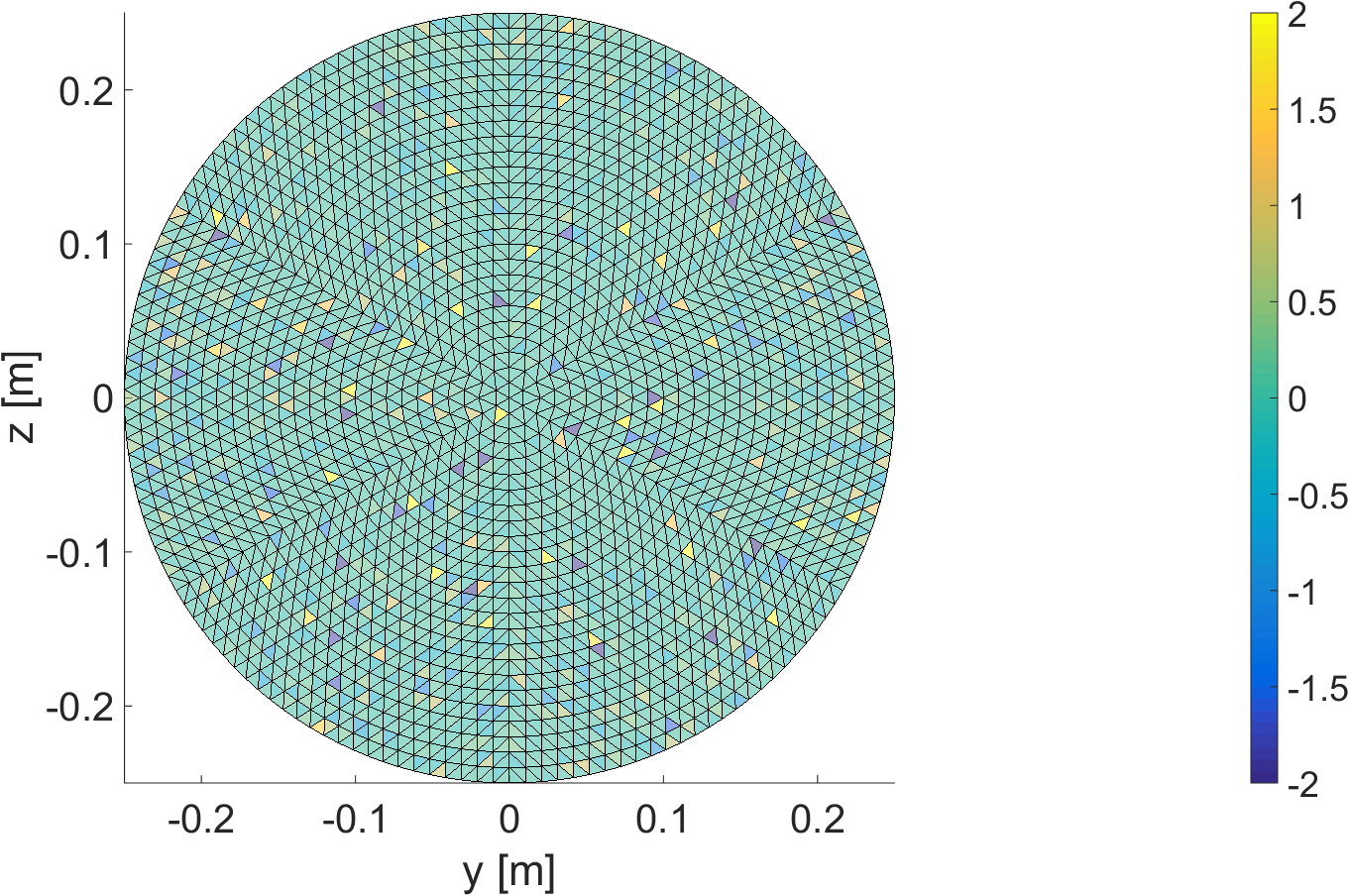}
    \caption{Difference in dielectric constant between the capacity with null optimized CRA and the baseline randomized CRA.}
   \label{fig:capacity_with_null_difference_result}
\end{figure}

\begin{figure}[h!]
\centering
\includegraphics[width=7.4cm, clip=true]{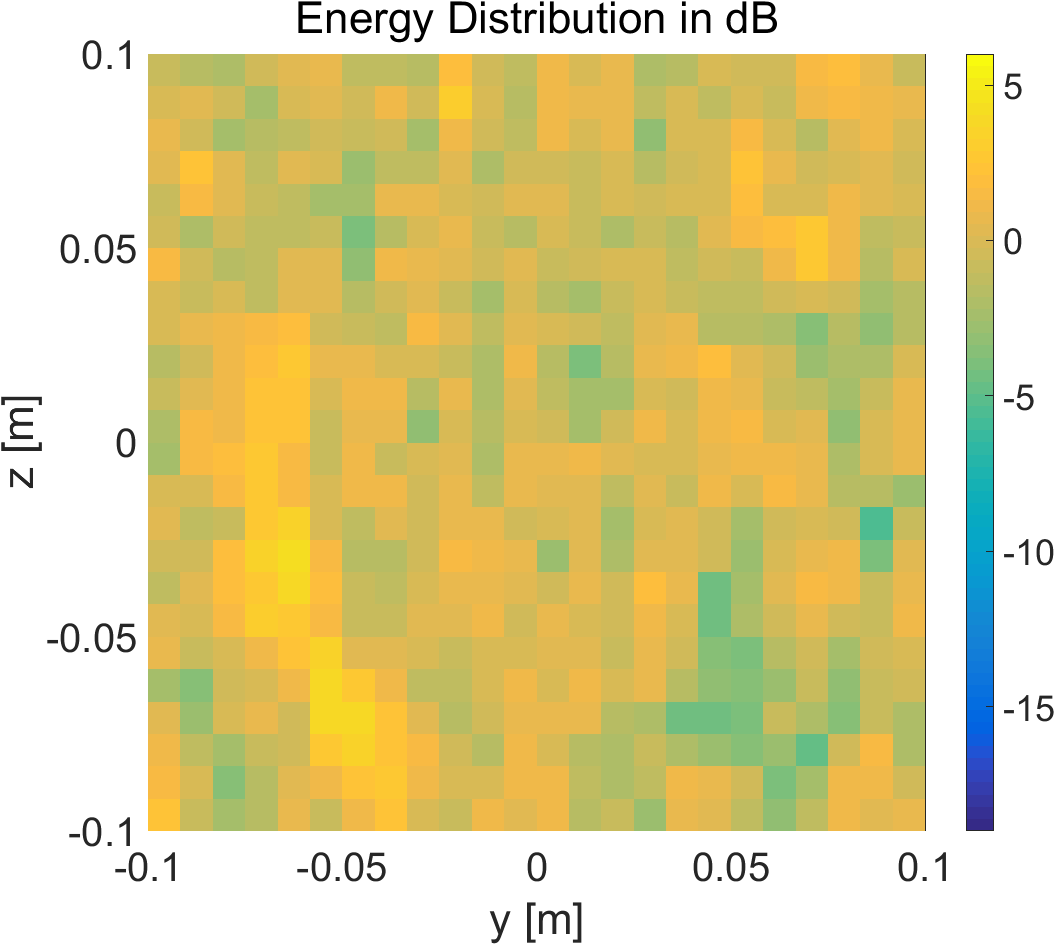}
    \caption{Energy distribution of the Green's functions of the baseline randomized CRA.}
   \label{fig:cra_random_receiver_gf_energy_dist}
\end{figure}

\begin{figure}[h!]
\centering
\includegraphics[width=7.4cm, clip=true]{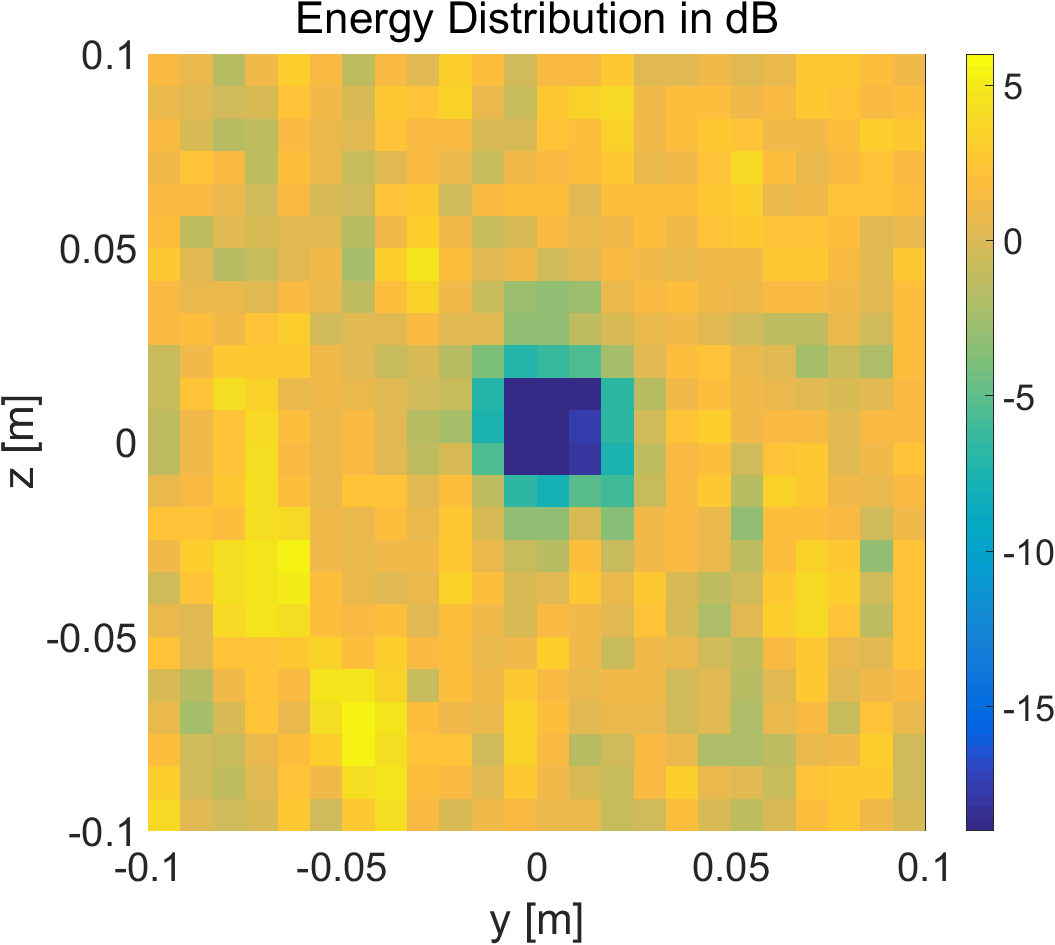}
    \caption{Energy distribution of the Green's functions of the capacity with null optimized CRA.}
   \label{fig:cra_optimized_receiver_gf_energy_dist}
\end{figure}

The design method has also been assessed for a communication application in which there is a strong interfering signal located near $x=0.9$[m], $y=z=0$[m]. The objective, then, was to optimize the capacity of the receiver Green's functions while simultaneously attenuating the interfering signal. This was achieved by solving Eq. \ref{eq:unified_opt} using the parameters displayed in Table \ref{tab:comm_design_tab}. Figure \ref{fig:capacity_with_null_difference_result} displays the difference in dielectric constant between baseline randomized CRA and the optimized CRA. Once again, the dielectric constants did not deviate significantly from the starting values. Nevertheless, the design algorithm produced a CRA that has an enhanced capacity ($-5747.1$ vs. $-5681.8$) and that significantly attenuates signals from the designated location. The latter property can be seen in Figures \ref{fig:cra_random_receiver_gf_energy_dist} and \ref{fig:cra_optimized_receiver_gf_energy_dist}, which display the energy distribution of the Green's functions for the baseline randomized CRA and the optimized CRA. 

\begin{table}[!h]
\captionsetup{font=normal}
\centering
\begin{tabular}{ |C{1.3cm}||C{5cm}|  }
 \hline
 \multicolumn{2}{|c|}{Design Parameters} \\
 \hline
& Receiver Capacity with Null. \\
 \hline
$\alpha_A$ & $0$ \\
\hline
$\alpha_{r_i}$ & $1$ \\
\hline
$\alpha_{t_i}$ & $0$ \\
\hline
$\beta_A$ & $10^{-6}$ \\
\hline
$\beta_{r_i}$ & $10^{-6}$ \\
\hline
$\beta_{t_i}$ & $10^{-6}$ \\
\hline
$\mbold{\Lambda}_{A,m}$ & $\mbold{0}_N$ \\
\hline
$\mbold{\Lambda}_{r,i,m}$ & Diagonal, all elements equal to $0$ except for points near $x=0.9$[m], $y=z=0$[m], which equal $-30$ \\
\hline
$\mbold{\Lambda}_{t,i,m}$ & $\mbold{0}_N$ \\
 \hline
\end{tabular}
\caption{Optimization parameters for communication design.}
\label{tab:comm_design_tab}
\end{table}

\vspace{7pt}
\section{Conclusion}
\label{sec:conclusion}
This paper describes a unified CRA design method for electromagnetic sensing and imaging applications. Unlike the previous design methods, the design method presented in this paper properly considers the coupling between the transmitting and receiving antennas. Furthermore, by including both sensing capacity and efficiency terms in the objective function, the unified design method allows one to design CRA's with improved CS reconstruction performance without significantly degrading the antenna's efficiency.

To demonstrate the capabilities of the unified method, a dielectric CRA was designed using the dielectric constants of the scatterers as the optimization variables and using MECA as the forward model. Although this design was shown to provide enhanced CS reconstruction performance, it admittedly may have little practical use, as it may be prohibitively difficult to manufacture such an antenna. Nevertheless, this paper provides the framework for designing more practical CRA's. In particular, it should be straightforward to extend the design method to optimize the thicknesses of the scatterers, which all have the same dielectric constant. This method is more amenable to practical imaging applications, as these CRA's can be manufactured efficiently using a $3$D printer. This method will be investigated in future work.

\vspace{7pt}

\bibliographystyle{IEEEtran}
\bibliography{./SICA-TA}

\end{document}